\documentclass[fleqn,usenatbib]{mnras}

\usepackage[T1]{fontenc}
\usepackage{ae,aecompl}

\usepackage{eqnarray}
\usepackage{graphicx}	
\usepackage{amsmath}	
\usepackage{amssymb}	

\title[A candidate kilonova in GRB160821B]{
The afterglow and kilonova of the short GRB 160821B}

\author[Troja et al.]{E. Troja$^{1,2}$\thanks{E-mail: eleonora@umd.edu}, 
A. J. Castro-Tirado$^{3,4}$, 
J. Becerra Gonz\'alez$^{5,6}$,
Y. Hu$^{3,7}$,
G. S. Ryan$^{1,8}$,
\newauthor
S. B. Cenko$^{2,1,8}$,
R. Ricci$^{9}$,
G. Novara$^{10}$,
R. S\'anchez-R\'amirez$^{11}$,
J. A. Acosta-Pulido$^{6,5}$,
\newauthor
K.~D.~Ackley$^{12}$
M. D. Caballero Garc\'ia$^{13}$,
S. S. Eikenberry$^{12}$,
S. Guziy$^{14}$,
S. Jeong$^{15}$,
\newauthor
A. Y. Lien$^{2}$, 
I. M\'arquez$^{3}$,
S. B. Pandey$^{16}$,
I.~H.~Park$^{15}$,
T. Sakamoto$^{17}$,
J. C. Tello$^{13}$,
\newauthor
I. V. Sokolov$^{18}$,
V. V. Sokolov$^{19}$,
A. Tiengo$^{9,20,21}$,
A. F. Valeev$^{19}$,
B. B. Zhang$^{22,23}$, 
\newauthor
S. Veilleux$^{1,8}$
\\
$^{1}$ Department of Astronomy, University of Maryland, College Park, MD 20742-4111, USA \\
$^{2}$ Astrophysics Science Division, NASA Goddard Space Flight Center, 8800 Greenbelt Rd, Greenbelt, MD 20771, USA\\
$^{3}$ Instituto de Astrof\'isica de Andalucia (IAA-CSIC), Glorieta de la Astronomia s\/n, E-18008, Granada, Spain\\
$^{4}$ Unidad Asociada Departamento de Ingenier\'ia de Sistemas y Autom\'atica, E.T.S. de Ingenieros Industriales, Universidad de M\'alaga, Spain\\
$^{5}$  Universidad de La Laguna, Dpto. Astrof\'isica, E-38206 La Laguna, Tenerife, Spain\\
$^{6}$ Instituto de Astrof\'isica de Canarias, E-38200 La Laguna, Tenerife, Spain\\
$^{7}$Universidad de Granada, Facultad de Ciencias Campus Fuentenueva S/N CP 18071 Granada, Spain\\
$^{8}$ Joint Space Science Institute, University of Maryland, College Park, MD 20742-4111, USA\\
$^{9}$ INAF-Istituto di Radioastronomia, Via Gobetti 101, I-40129, Italy\\
$^{10}$ Scuola Universitaria Superiore IUSS Pavia, Piazza della Vittoria 15, I-27100 Pavia, Italy\\
$^{11}$ INAF, Istituto di Astrofisica e Planetologia Spaziali, via Fosso del Cavaliere 100, 00133 Rome, Italy\\
$^{12}$ Department of Astronomy, University of Florida, Gainesville, FL 32611\\
$^{13}$ Astronomical Institute of the Academy of Sciences, Bo\v{c}n\'{\i}~II~1401, CZ-14100 Praha 4, Czech Republic\\
$^{14}$  Nikolaev National University, Nikolska 24, Nikolaev 54030, Ukraine\\
$^{15}$ Institute for Science and Technology in Space, SungKyunKwan University, Suwon 16419, Korea\\
$^{16}$ Aryabhatta Research Institute of Observational Sciences, Manora Peak, Nainital - 263 002, India\\
$^{17}$ Department of Physics and Mathematics, Aoyama Gakuin University, 5-10-1 Fuchinobe, Chuoku, Sagamiharashi Kanagawa 252-5258, Japan \\
$^{18}$ Terskol Branch of Institute of Astronomy of the RAS, 48 Pyatnitskaya st., Moscow, 119017 Russia\\
$^{19}$ Special Astrophysical Observatory of RAS, Nizhniy Arkhyz, Zelenchukskiy region, Karachai-Cherkessian Republic, 369167, Russia\\
$^{20}$ INAF, Istituto di Astrofisica Spaziale e Fisica Cosmica Milano, Via E. Bassini 15, I-20133 Milano, Italy\\ 
$^{21}$ INFN-Istituto Nazionale di Fisica Nucleare, Sezione di Pavia, via A. Bassi 6, I-27100 Pavia, Italy\\
$^{22}$ School of Astronomy and Space Science, Nanjing University, 163 Xian-Lin Ave., Nanjing 210023, China\\
$^{23}$ Key Laboratory of Modern Astronomy and Astrophysics (Nanjing University), Ministry of Education, China\\
}
\date{Accepted 2019 Aug 12. Received 2019 Aug 12; in original form 2019 May 3}

\pubyear{2019}

\begin{document}

\pagerange{\pageref{firstpage}--\pageref{lastpage}}
\maketitle
\label{firstpage}

\begin{abstract}

GRB~160821B is a short duration gamma-ray burst (GRB) detected and localized by the {\it Neil Gehrels Swift Observatory} in the outskirts of a spiral galaxy at $z$=0.1613, at a projected physical offset of $\approx$16~kpc from the galaxy's center.
We present X-ray, optical/nIR and radio observations of its counterpart and model them with two distinct components of emission:
a standard afterglow, arising from the interaction of the relativistic jet with the surrounding medium, 
and a kilonova, powered by the radioactive decay of the sub-relativistic ejecta. 
Broadband modeling of the afterglow data reveals a weak reverse shock propagating backward into the jet, 
and a likely jet-break at $\approx$3.5 d. This is consistent with a 
structured jet seen slightly off-axis ($\theta_{view} \sim \theta_{core}$) while expanding into a low-density medium ($n \approx$10$^{-3}$\,cm$^{-3}$). 
Analysis of the kilonova properties suggests a rapid evolution toward red colors, similar to AT2017gfo, and a low nIR luminosity, possibly due to the presence of a long-lived neutron star.
The global properties of the environment, the inferred low mass ($M_{ej} \lesssim$ 0.006 $M_{\odot}$)
and velocities ($v_{ej} \gtrsim$0.05$c$) of lanthanide-rich ejecta are consistent with a
binary neutron star merger progenitor.

\end{abstract}



\section{Introduction}

Short duration GRBs were long suspected to be the product of compact binary mergers \citep{Blinnikov84,Goodman86,Paczynski86,Eichler89,Narayan92}, 
involving either two neutron stars (NSs) or a NS and a solar-mass black hole (BH). 
The merger remnant, either a massive NS \citep{Bucciantini11,Giacomazzo13} or an accreting BH \citep{Ruffert99,Baiotti08,Kiuchi09}, 
launches a highly relativistic jet, which produces the GRB emission  \citep{Rezzolla11, Paschalidis15, Ruiz16}. Later, the interaction of this jet with the surrounding medium produces the observed broadband (from radio to X-rays) afterglow spectrum via synchrotron radiation \citep[e.g.][]{Granot02}.
Within this framework, the study of the GRB afterglow probes the jet structure and geometry as well as the properties of the surrounding environment. 


Another long-standing prediction of the NS merger model is the presence of a luminous, short-lived transient arising from the radioactive decay of freshly synthesized r-process elements \citep{Li98,Metzger10,Barnes13,Tanaka13}. Whereas the afterglow does not provide any direct link to the GRB progenitor, such radioactively powered transient, initially dubbed  ``mini-supernova'' and now more commonly referred to as ``kilonova'', is the distinctive signature of compact binary mergers and clear signpost of heavy elements production in these systems \citep{Lattimer74,Lattimer76, Freiburghaus99,Roberts11,Goriely11, Korobkin12,Grossman14,Rosswog14}. 
Evidence of kilonovae associated to short GRBs was only recently found \citep{Tanvir13,Yang15, Jin16}. The timescales and red color of these first candidate kilonovae suggested that the merger ejecta were highly opaque, as predicted for the heaviest r-process elements \citep{Barnes13,Tanaka13}. The high opacity causes any UV/optical emission to be significantly suppressed, thus naturally explaining the lack of kilonova detections in over a decade of {\it Swift} observations \citep{Bloom06,Kocevski10,Kann11,Pandey19}. 

This slowly-progressing field was revolutionized by the discovery of GW170817 and its electromagnetic counterparts
\citep{Abbott17a,Abbott17b}. 
X-ray observations revealed the onset of an off-axis afterglow \citep{Troja17} powered by a relativistic structured jet \citep{Mooley2018,Ghirlanda19,Troja19}.
Thanks to the proximity of the event, the associated kilonova AT2017gfo was characterized in great detail \citep[e.g.][]{Covino17,Pian17,Smartt17,Tanvir17,Troja17}.
In addition to the expected red emission, peaking in the nIR a few days after the merger, a distinctive feature of AT2017gfo was its luminous UV/optical light, peaking at early times and rapidly fading away \citep{Evans17}. 
Although the origin of this blue component remains an open question, it immediately revealed a complex chemical
composition and velocity structure of the merger ejecta, indicating that the phenomenology of a kilonova is likely determined by the interplay of multiple outflows emerging from the merger remnant \citep{Kasen17,Perego17,Radice18,Wollaeger18}.

Fostered by the discovery of AT2017gfo and its luminous blue emission, several attempts were made to find similar cases in archival short GRB observations \citep{Gompertz18,Troja18b,Rossi19}.
\cite{Troja18b} found that some nearby events
have optical luminosities comparable to AT2017gfo. In particular, they showed that the short GRB150101B was a likely analogue to GW170817, characterized by a late-peaking afterglow and a luminous optical kilonova emission, dominating at early times. 
Other, although less clear, cases discussed in the literature include GRB060505 \citep{Ofek07} and GRB080503 \citep{Perley09}.  
Overall, it seems plausible that kilonovae similar to AT2017gfo could have been detected in the optical, although not clearly identified prior to GW170817. 

Whereas this observational evidence suggests that r-process nucleosynthesis
is common in the aftermath of a short GRB, it does not inform 
on the production of the heaviest elements, i.e. those with atomic mass number $A$\,$\gtrsim$140.
NS merger ejecta with an electron fraction $Y_e$\,$\gtrsim$0.25 
do not have enough neutrons to push the nuclear chain past the second r-process peak at $A$\,$\approx$130, thus producing only
a blue and fast-fading kilonova \citep{Li98,Metzger10}. 
An efficient production of lanthanides and actinides will increase the ejecta opacity, leading to a delayed and redder kilonova emission \citep{Barnes13,Tanaka13}. Only the detection of such red component, peaking at IR wavelengths, is a clear signature of
the production of heavy ($A$\,$\gtrsim$140) r-process nuclei. 
Unfortunately, archival IR observations of short GRBs are sparse and  mostly unconstraining, and the only possible nIR detection of a kilonova in a short GRB remains GRB130603B \citep{Tanvir13}.

The short duration GRB160821B, thanks to its low redshift and a rich multi-wavelength dataset, is an excellent 
testbed for kilonova searches and afterglow studies. 
The presence of a kilonova was discussed by \citet{Kasliwal17} and \citet{Jin18}, who did not find conclusive evidence
for it due to the sparse dataset used in both studies. 
Here we present a comprehensive broadband analysis of this event, including data from {\it Swift} and {\it XMM-Newton} in the X-rays,
the Gran Telescopio de Canarias (GTC), the William Herschel Telescope (WHT), the {\it Hubble Space Telescope} (HST), the Keck I telescope,  and the Discovery Channel Telescope (DCT) in the optical/nIR, 
and the Jansky Very Large Array (VLA) in the radio. 
Thanks to the good temporal and spectral coverage of our dataset, we resolve two emission components, which we identify as the GRB afterglow and its associated kilonova. We can exclude dust as the origin of the observed red color, and interpret it as evidence for a lanthanide-rich kilonova emission. 

Throughout the paper, we adopt a standard $\Lambda$CDM cosmology \citep{Planck2018}. Unless otherwise stated, the quoted errors are at the 68\% confidence level, and upper limits are at the 3~$\sigma$ confidence level.

\section{Observations and Data Analysis}

GRB 160821B triggered the {\it Swift} Burst Alert Telescope (BAT; \citealt{Barthelmy05}) at 22:29:13 UT on 2016 August 21, hereafter referred to as $T_0$.   Based on its duration of $T_{90}$\,=0.48$\pm$0.07~s \citep{Lien16}, it is classified as a short burst.
According to the BAT GRB catalogue\footnote{https://swift.gsfc.nasa.gov/results/batgrbcat/}, 
its spectrum is relatively soft, and best described ($\chi^2$=59 for 59 dof) by a power-law with a high-energy exponential cut-off at $\approx$50~keV. The time-averaged fluence is $(1.10 \pm 0.10) \times 10^{-7}$\,erg\,cm$^{-2}$\,s$^{-1}$
in the 15-150 keV energy band. 
The Fermi GBM Burst catalogue\footnote{https://heasarc.gsfc.nasa.gov/W3Browse/fermi/fermigbrst.html} reports
a fluence
of $(2.0 \pm 0.2) \times 10^{-7}$\,erg\,cm$^{-2}$\,s$^{-1}$
over the broader the 10-1,000 keV energy band.
Using this value and the redshift $z$\,$\sim$\,0.1613 from the candidate host galaxy (see Sect. 2.2.6), we derive an isotropic-equivalent energy $E_{\gamma,iso}$\,=\,(1.3 $\pm$ 0.3) $\times$\,$10^{49}$\,erg in the 10-1,000 keV energy band,
at the lower end of the observed distribution of short GRB energetics. 

Its short duration and possible low redshift made it a prime target
for kilonova searches, triggering an intense multi-wavelength campaign, as we describe in detail below. 
A log of the observations is reported in Table~\ref{tab:log}.

\subsection{X-rays}

\subsubsection{Swift/XRT}\label{sec:xrt}
Observations with the X-Ray Telescope (XRT; \citealt{Burrows05})
on-board {\it Swift} started 57~s after the trigger, and monitored the afterglow for the following 23 days for a total net exposure of 198 s in Windowed Timing (WT) mode and 39 ks in Photon Counting (PC) mode. {\it Swift} data were retrieved from the public on-line repository\footnote{http://www.swift.ac.uk/xrt\_products/} \citep{Evans09}
by using custom options for the light curve binning and standard settings for the spectral extraction. 
Spectra were modelled within XSPEC v.12.10.1 \citep{Arnaud96} by minimizing the Cash statistics \citep{Cash79}.
The Galactic absorption column was modelled with the function \texttt{tbabs} and fixed to the value $N_{\rm H}=5.7\times10^{20}$\,cm$^{-2}$ \citep{Willingale13}.
The time-averaged spectrum, from 4~ks to 74~ks, is well described by
an absorbed power-law function with photon index ${\Gamma=1.8\pm0.2}$. 
Based on this model, we derive a counts-to-unabsorbed flux conversion factor of 4.4$\times$10$^{-11}$ erg cm$^{-2}$ ct$^{-1}$. The derived X-ray fluxes are reported in Table~\ref{tab:log}.
No evidence for intrinsic absorption was found at the galaxy's redshift of
$z$$\sim$0.1613, and we place a 3\,$\sigma$~upper limit 
${N_{{\rm H},z} < 5.5\times10^{21}{\text{\,cm}^{-2}}}$ using the model \texttt{ztbabs}.

\subsubsection{XMM-Newton} 
The X-ray afterglow was also observed with {\it XMM-Newton} at two epochs, on 2016 August 25 (obsID: 0784460301) and on 2016 August 31 (obsID: 0784460401; PI: Tanvir). The PN \citep{Struder01} and two MOS \citep{Turner01} CCD cameras operated in Full Frame mode and with the thin optical blocking filter. The PN, MOS1 and MOS2 exposure times were, respectively: 24.6 ks, 27.1 ks, 27.1 ks for the first observation and 34.1 ks, 36.0 ks, 35.9 ks for the second observation.

Data were retrieved from the public archive,
and processed using the XMM-Newton Science Analysis Software (SAS; \citealt{Gabriel04}) version 16.1.0. After removing intervals of high particle background, the effective exposures were reduced by $\approx$55\% for the first observation and $\approx$30\% for the second observation.
For our analysis we selected only events with FLAG=0 and PATTERN$\leq$4 and PATTERN$\leq$12 for PN and MOS, respectively. 

The afterglow is detected with high significance during the first {\it XMM} epoch. Source spectra were extracted from a circular region with radius of 20\,\arcsec, and the background was estimated from two independent source-free boxes around the afterglow position. 
We extracted 57 source counts from the PN exposure, and modeled the spectrum as described in Sect.~\ref{sec:xrt} with an absorbed power-law function of photon index ${\Gamma=1.88\pm0.24}$. 
During the second epoch, the source is still visible with a detection likelihood (DET\_ML=7.9) slightly above the standard detection threshold adopted for the {\it XMM-Newton} source catalogues (DET\_ML=6). 
 By summing PN and MOS data, we extract a total of 47 source counts. The corresponding X-ray flux was derived by assuming the same spectral parameters of the first epoch.
Our results are reported in Table~\ref{tab:log}.

\begin{table*}
        \centering
        \caption{Observations of GRB160821B.}
        \label{tab:log}
        \begin{tabular}{cccccccc}
        \hline
        Date & T-T$_0$  & 
        Telescope & Instrument & 
        Exposure  & Band & 
        AB mag & Flux density\footnotemark   \\
         MJD & (d) &
            &     &
         (s) &    & 
             &  ($\mu$Jy)  \\
        \hline
       \multicolumn{8}{c}{Optical/nIR}\\
      \hline
       57621.940 & 0.002 & {\it Swift} & UVOT & 147 & $wh$ & $>$21.9 & $<$8 \\
       57621.942 & 0.004 & {\it Swift} & UVOT & 209 & $u$ & $>$21.4 &  $<$12\\
       57622.013 & 0.076 & GTC & OSIRIS & 270 & $r$ & 22.67$\pm$0.10 & 3.4$\pm$0.3 \\
       57622.017 & 0.080 & GTC & OSIRIS & 270 & $i$ & 22.39$\pm$0.07 & 4.3$\pm$0.3 \\
       57622.020 & 0.083 & GTC & OSIRIS & 180 & $z$ & 22.28$\pm$0.06 & 4.7$\pm$0.3 \\
       57623.006 & 1.07 & WHT & ACAM & 1440 & $r$ & 23.83$\pm$0.25 & 1.2$\pm$0.3 \\
       57623.027 & 1.09 & WHT & ACAM & 1680 & $z$ & 23.6$\pm$0.3 & 1.4$\pm$0.4 \\
       57623.878 & 1.941 & GTC & CIRCE  &   540 &   $H$    & $>$23.8  & $<$1.1\\
       57623.895 & 1.958 & GTC & CIRCE & 1800 &   $J$   &   $>$24.0 &  $<$0.9\\
       57623.921 & 1.984 & GTC & CIRCE &  600 &  $K_s$  & $>$23.3 & $<$1.7\\

       57623.958 & 2.021 & GTC & OSIRIS & 800 & $g$ & 25.67$\pm$0.15  & 0.22$\pm$0.03 \\
       57623.965 & 2.028 & GTC & OSIRIS & 720 & $r$ & 25.12$\pm$0.12  & 0.36$\pm$0.04 \\
       57623.973 & 2.036 & GTC & OSIRIS & 450 & $i$ & 24.56$\pm$0.12  & 0.58$\pm$0.06 \\
       57623.980 & 2.043 & GTC & OSIRIS & 420 & $z$ & 24.31$\pm$0.17  & 0.72$\pm$0.11 \\
       57625.564 & 3.627 & HST & WFC3 & 2484 & $F606W$ & 26.02$\pm$0.06 & 0.157$\pm$0.009 \\
       57625.631 & 3.694 & HST & WFC3 & 2397 & $F160W$ & 24.53$\pm$0.08  & 0.57$\pm$0.04 \\
       57625.697 & 3.760 & HST & WFC3 & 2397 & $F110W$ & 24.82$\pm$0.05  & 0.44$\pm$0.02 \\
      57625.929 & 3.992 & GTC & OSIRIS & 450 & $g$ & $>$25.6  & $<$0.24 \\
      57625.934 & 3.997 & GTC & OSIRIS & 420 & $i$ & $>$25.8  & $<$0.19 \\
      57626.234 & 4.297 & Keck I & MOSFIRE & 160 & $K_s$ & 24.0$\pm$0.4 &  0.9$\pm$0.3\\
      57626.922 & 4.985 & GTC & OSIRIS & 800 & $r$ & 26.49$\pm$0.20  & 0.101$\pm$0.019 \\
      57629.402 & 7.465 & Keck I & MOSFIRE & 145 & $K_s$ & $>$23.9 &  $<$0.9\\
      57630.321 & 8.383 & Keck I & MOSFIRE & 110 & $K_s$ & $>$23.7 &  $<$1.2\\
      57631.924 & 9.987 & GTC & OSIRIS & 720 & $i$ & $>$26.0 &  $<$0.15 \\
      57631.935 & 9.998 & GTC & OSIRIS & 1200 & $g$ & $>$25.8 &  $<$0.20 \\
      57631.950 & 10.013 & GTC & OSIRIS & 960 & $r$ & $>$26.2 & $<$0.13 \\
      57632.325 & 10.388 & HST & WFC3 & 1863 & $F606W$ & 27.9$\pm$0.3  & 0.028$\pm$0.008 \\
      57632.383 & 10.446 & HST & WFC3 & 2397 & $F110W$ & 26.9$\pm$0.4  & 0.07$\pm$0.02 \\
      57632.449 & 10.512 & HST & WFC3 & 2397 & $F160W$ & 26.6$\pm$0.3  & 0.08$\pm$0.02 \\
      57645.088 & 23.151 & HST & WFC3 & 1350 & $F606W$ & $>$27.2  & $<$0.05 \\
      57645.108 & 23.171 & HST & WFC3 & 1497 & $F110W$ & $>$26.6  & $<$0.09 \\
      57645.154 & 23.217 & HST & WFC3 & 2097 & $F160W$ & $>$25.7  & $<$0.19 \\ 
      57721.2 &  99.2 & HST & WFC3 & 5395 & $F110W$ & reference  & -- \\ 
      57725.3 & 103.3 & HST & WFC3 & 2484 & $F606W$ & reference  & -- \\
      58333.7 & 711.7 & HST & WFC3 & 2796 & $F160W$ & reference  & -- \\

      \hline
        \multicolumn{8}{c}{Radio}\\
      \hline
      57622.11 & 0.17  & VLA & -- & 3600 & C & -- & 26 $\pm$ 5 \\
      57623.06 & 1.13  & VLA & -- & 3600 & C & -- & $< 15$     \\
      57632.01 & 10.07 & VLA & -- & 6480 & X & -- & $< 11$     \\
      57639.04 & 17.10 & VLA & -- & 6460 & X & -- & $<33$      \\ 
      \hline
      \multicolumn{8}{c}{X-ray}\\
      \hline 
      57721.995 & 0.057 & {\it Swift} & XRT & 185 & 0.3 -- 10 keV & -- & 0.15$^{+0.08}_{-0.06}$ \\
      57722.001 & 0.063 & {\it Swift} & XRT & 566 & 0.3 -- 10 keV & -- & 0.05$^{+0.03}_{-0.02}$ \\
      57722.008 & 0.070 & {\it Swift} & XRT & 784 & 0.3 -- 10 keV & -- & 0.05$\pm$0.02 \\
      57722.064 & 0.126 & {\it Swift} & XRT & 363 & 0.3 -- 10 keV & -- & 0.08$^{+0.04}_{-0.03}$ \\
      57722.068 & 0.130 & {\it Swift} & XRT & 396 & 0.3 -- 10 keV & -- & 0.07$^{+0.04}_{-0.03}$ \\
      57722.074 & 0.136 & {\it Swift} & XRT & 865 & 0.3 -- 10 keV & -- & 0.037$^{+0.018}_{-0.014}$ \\
      57722.132 & 0.195 & {\it Swift} & XRT & 862 & 0.3 -- 10 keV & -- & 0.035$\pm$0.019 \\
      57722.222 & 0.285 & {\it Swift} & XRT & 1273 & 0.3 -- 10 keV & -- & 0.025$^{+0.014}_{-0.010}$ \\
      57722.264 & 0.327 & {\it Swift} & XRT & 584 & 0.3 -- 10 keV & -- & 0.05$\pm$0.02 \\
      57722.278 & 0.340 & {\it Swift} & XRT & 1584 & 0.3 -- 10 keV & -- & 0.029$\pm$0.011 \\
      57722.356 & 0.419 & {\it Swift} & XRT & 3908 & 0.3 -- 10 keV & -- & 
      0.014$\pm$0.05 \\
      57722.962 & 1.024 & {\it Swift} & XRT & 9008 & 0.3 -- 10 keV & -- & 
      (3.6$\pm$0.2)$\times$10$^{-3}$  \\
      57724.725 & 2.327 & {\it Swift} & XRT & 8777 & 0.3 -- 10 keV & -- & 
      $<$4.6$\times$10$^{-3}$  \\
       57625.879 & 3.942 & XMM-Newton & EPIC/PN & 10880 & 0.3 -- 10 keV & -- & (2.3$\pm$0.3)$\times$10$^{-3}$  \\
       57631.913 & 9.976 & XMM-Newton & EPIC/PN & 22665 & 0.3 -- 10 keV & -- & (2.9$\pm$1.5)$\times$10$^{-4}$ \\
      57737.570 & 15.172 & {\it Swift} & XRT & 26000 & 0.3 -- 10 keV & -- &  $<$10$^{-3}$  \\
      \hline
    \end{tabular}
\begin{flushleft}    
{$^3$ \footnotesize Optical/nIR fluxes were corrected for Galactic extinction due to the reddening $E(B-V)=$0.04 along the sightline
\citep{Green18}. 
X-ray fluxes were corrected for Galactic absorption 
$N_H$=5.7$\times$10$^{20}$\,cm$^{-2}$ \citep{Willingale13}, and converted into
flux densities at 1 keV using the best fit photon index $\Gamma$=1.88.}
\end{flushleft}
\end{table*}

\subsection{Optical/nIR}

\subsubsection{Swift/UVOT} 
Observations with the Ultra-Violet Optical Telescope (UVOT; \citealt{Roming06}) on-board {\it Swift} started 76~s after the trigger. The GRB position was initially imaged with the
$white$ and $u$ filters, but no counterpart was detected. 
We used the zero-points provided by \citet{Breeveld11} to convert UVOT count rates into the AB magnitude system \citep{Oke74}.
The corresponding 3 $\sigma$ upper limits are reported in Table~\ref{tab:log}.
Subsequent observations used all the optical and UV filters,
and are reported in \citet{Breeveld16}.

\subsubsection{Gran Telescopio Canarias (GTC)} 

We observed the GRB afterglow (PI: A. Castro-Tirado) with the 10.4 m Gran Telescopio de Canarias (GTC), located at the observatory of Roque de los Muchachos in La Palma (Canary Islands, Spain), equipped with the Optical System for Imaging and low-intermediate-Resolution Integrated Spectroscopy (OSIRIS) 
and the Canarias InfraRed Camera Experiment (CIRCE) instruments. 
Deep optical images in the $g$, $r$, $i$, and $z$ filters were taken over five different nights, starting as early as 1.8~hr after the trigger, in order to characterize any spectral evolution of the GRB counterpart
 \citep{Xu2016}.
A single epoch of nIR imaging in the $J$, $H$, and $K_s$ filters was 
carried out $\approx$2 d after the trigger.

Data were reduced and aligned in a standard fashion using a custom pipeline, based mainly on {\it Astropy} and {\it photutils} python libraries. 
We combined the images by weighting each individual frame based on its depth, and applying a 3$\sigma$ clipping algorithm. Photometric zero points and astrometric calibration were computed using the Pan-STARRS catalogue \citep{Chambers16}.
We then performed point spread function (PSF) matching photometry of the afterglow on the combined images. To construct the empirical PSF, we selected bright and isolated stars close to the afterglow, and combined them weighting by their flux.  Background and host light contamination were left as free parameters in the fit.
The resulting values are reported in Table~\ref{tab:log}.

On August 23 ($\approx$\,$T_0$+2\,d), a  GTC (+OSIRIS) spectrum (3 $\times$ 1500s) with the R1000B grism and a 1.\arcsec slit covering the 3,700,\AA-7,500, \AA~range was gathered with the slit being placed in order to cover the
GRB location. Data were reduced and calibrated using standard routines. No absorption or emission lines can be detected superimposed on the continuum emission.

\subsubsection{William Herschel Telescope (WHT)}

Optical imaging with the Auxiliary port CAMera (ACAM) started on 2016-08-22 at 23:55
($\approx T_0$ + 1.1~d; \citealt{Levan16}). Observations were carried out in the $r$ filter for a total integration of 24~min and in $z$ filter for a total exposure of 28~min. Seeing during the observations was around 0.8$^{\prime\prime}$. 
The data were reduced within IRAF\footnote{IRAF is distributed by the National Optical Astronomy Observatory, which is operated by the Association of Universities for Research in Astronomy (AURA) under cooperative agreement with the National Science Foundation.} following standard procedures (e.g., bias subtraction, flat-fielding, etc.). 
Aperture photometry was performed using the {\it photutils\footnote{\url{https://photutils.readthedocs.io/en/stable/}}} package, and calibrated to nearby Pan-STARRS sources \citep{Chambers16}. A 20\% systematic uncertainty was added to our measurements to account for contaminating light from the nearby  candidate host galaxy. Our measurements are reported in Table~\ref{tab:log}.

\subsubsection{Hubble Space Telescope (HST)}\label{hst}

We activated our program to search for kilonovae (GO14087,GO14607; PI: Troja) and, starting on August 22, obtained several epochs of deep imaging with the IR and UVIS channels of the Wide Field Camera 3 (WFC3).  A complete log of the observations is reported in Table~\ref{tab:log}. Data were processed through the STSCI pipeline, and standard tools within the
\textit{stsci\_python} package on AstroConda\footnote{AstroConda is a free Conda channel maintained by the Space Telescope Science Institute (STScI). It provides tools and utilities required to process and analyse data from the Hubble Space Telescope (HST), James Webb Space Telescope (JWST), and others: http://AstroConda.readthedocs.io/} were used  to align, drizzle and combine the exposures into the final images.  The resulting plate scales were 0.067\arcsec/pixel for the WFC3/IR images and  0.033\arcsec/pixel for the WFC3 F606W images.
Late-time observations of the host galaxy were used as reference templates. Aperture photometry was performed on the subtracted images, and the tabulated zero points were used to to determine the source brightness. 

\begin{figure*}
\includegraphics[width=1.9\columnwidth]{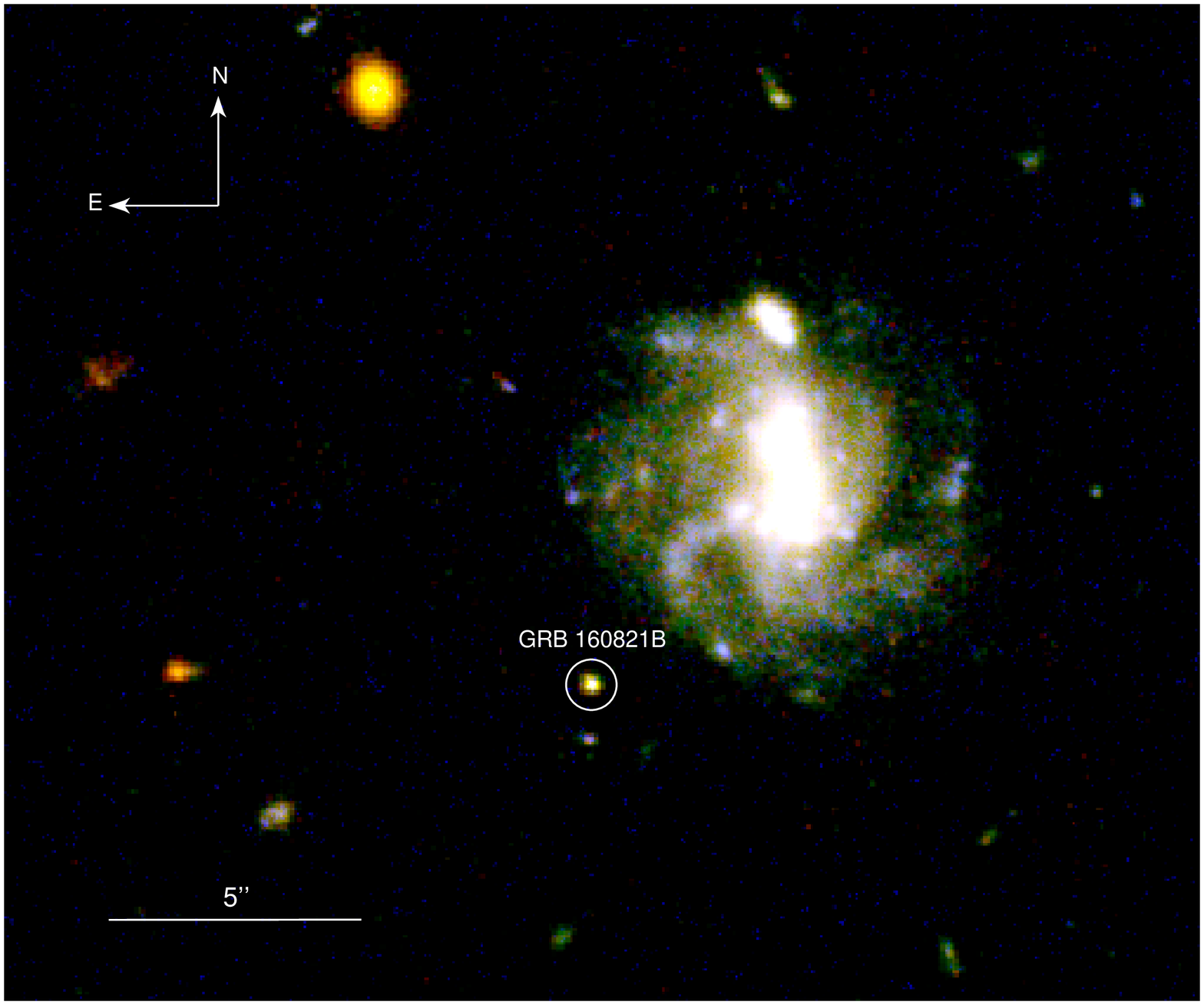}\vspace{0.4cm}
\includegraphics[width=1.85\columnwidth]{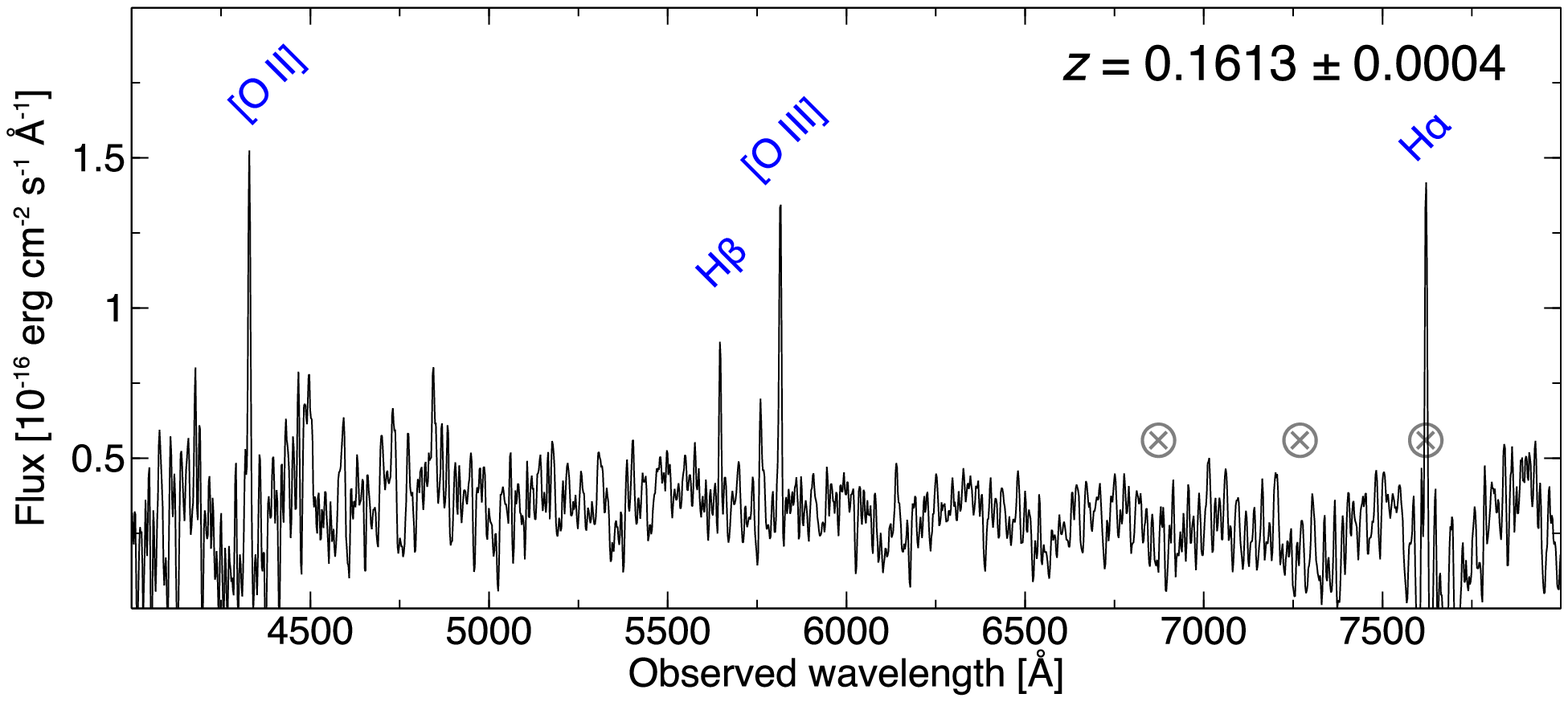}
     \caption{{\it Top panel:} the field of GRB160821B observed with HST using the 
     $F606W$ (blue), $F110W$ (green), and $F160W$ (red) filters. The white circle marks the position of the GRB counterpart, located $\approx$16 kpc from the center of its galaxy.
     {\it Bottom panel:} optical spectrum of the GRB host galaxy taken with the DeVeny spectrograph on the 4.3m DCT.
     The brightest lines are identified. The crossed circles mark the position of strong telluric features.}
     \label{fig:one}
     \vspace{-0.3cm}
\end{figure*}

The GRB counterpart is detected in all filters in the earlier two epochs, whereas is no longer visible in subsequent visits.  Its position, 
determined from the optical frames, is
R.A. (J2000) = 18:39:54.56, Dec (J2000) = +62:23:30.35 with a 1\,$\sigma$ uncertainty of 0.04\arcsec, and lies 5.7\arcsec~from the center of a bright face-on spiral (Figure~\ref{fig:one}, which is likely the GRB host galaxy. At a redshift z$\sim$0.1613, this corresponds to a projected physical offset of 16.40$\pm$0.12 kpc.

\subsubsection{Keck } 

 NIR observations with the MOSFIRE instrument on the Keck I telescope were taken at three different epochs,
as previously reported by \cite{Kasliwal17}. A possible detection was found during the first epoch at $T_0$+4.3~d, whereas the source was undetected at later times. 
We retrieved the archival data, and independently analyzed them using standard procedures for CCD data reduction. 
For the first epoch, four frames considered particularly noisy were removed, whereas the remaining ones were 
aligned using SCAMP \citep{Bertin06} and stacked with SWarp \citep{Bertin02}  into a final image with 160~s of total exposure. Our analysis confirms the presence of a weak signal ($\approx$3.5\,$\sigma$) at the GRB position. A magnitude of $K_s$=22.12$\pm$0.38 was derived by performing aperture photometry calibrated to nearby point sources
from the  Two Micron All Sky Survey (2MASS; \citealt{Skrutskie06}). Our result is in agreement with the value quoted by \citet{Kasliwal17}. We used the offsets from \citet{Blanton07} to convert the 2MASS Vega magnitudes to the AB system, as quoted
in Table~\ref{tab:log}.

\subsubsection{Discovery Channel Telescope (DCT)}

The DeVeny spectrograph on the 4.3-meter DCT was used on 2017 March 19 with the 300 g~mm$^{-1}$ grating in the first order and a 1.\arcsec5 slit to obtain a spectrum for the GRB host galaxy, covering the 3,600,\AA - 8,000\,\AA~range at a dispersion of $\sim$2.2\,\AA~per pixel. Standard IRAF procedures were used for reduction and calibration. 

The final spectrum is shown in Figure~\ref{fig:one} (bottom panel).
Several nebular emission lines are visible, including those at $\lambda_{obs}$ $\approx$ 4329, 5814, 5759, and 5647 \AA, associated with [O II], $H{\beta}$ and [O III] transitions, respectively, and identify the galaxy to be at $z$ = 0.1613$\pm$0.0004,  consistent with the preliminary estimate of \citet{Levan16}. The bright line at 7622 \AA, associated with the Balmer $H{\alpha}$ line, falls within the telluric A band. Lines properties were derived by modeling them with Gaussian functions using the {\it splot} task in IRAF.

\subsection{Radio}

Radio observations were carried out with the Karl J. Jansky Very Large Array (JVLA) at four different epochs, 
on Aug 22 and 23, 2016 (project code: 15A-235; PI: Fong), and on Sep 1 and 8, 2016 (project code: 16B-386; PI: Gompertz) in B configuration at the center frequencies of 6~GHz with a bandwidth of 2~GHz (the former two epochs) and 10~GHz with a bandwidth of 4~GHz (the latter two epochs). The primary calibrator was 3C286 and the phase calibrator was J1849+6705 for all the four epochs. The data were downloaded from the VLA Archive and calibrated using the JVLA CASA pipeline v1.3.11 running in CASA v4.7.2. The data were then split, imaged and cleaned using CASA in interactive mode: Briggs weighting, robustness = 0.5 and 1000 clean iterations were performed. The results are presented in Table~\ref{tab:log}. 
The afterglow was detected during the first epoch at a flux of $\approx$26~$\mu$Jy, consistent within the uncertainties with the value reported in \citet{Fong16}.
By performing a 2D Gaussian fit to the source region we obtain a position of R.A. (J2000) = 18:39:54.56, Dec (J2000) = 62:23:30.32
with an error of 0.07 arcsec and 0.08 arcsec, respectively.
This is consistent with the optical position (Sect.~\ref{hst}).
For the other epochs a $3~\sigma$ flux density upper limit is provided.


\section{Disentangling the Afterglow and Kilonova emission} \label{agkn}

\begin{figure}
\includegraphics[width=0.98\columnwidth]{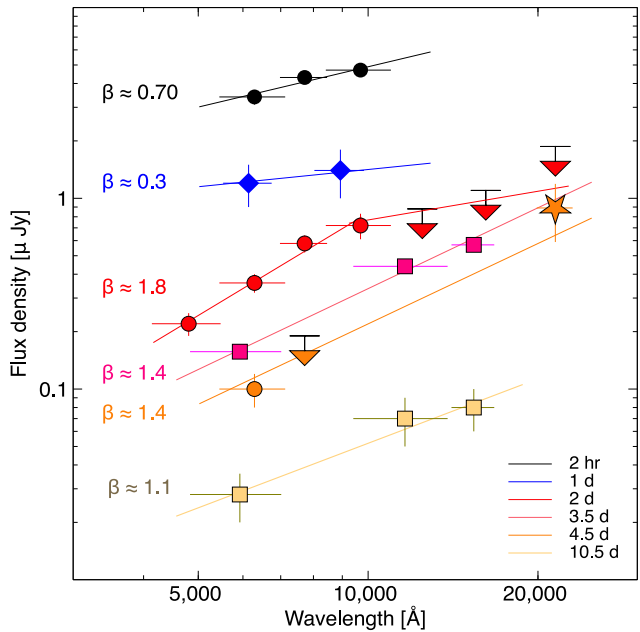}
     \caption{Color evolution of the optical/nIR counterpart, compiled including data from GTC (circles), WHT (diamonds), Keck (star), and HST (squares). }
     \label{fig:color}
     \vspace{-0.3cm}
\end{figure}

The kilonova AT2017gfo  was characterized by a quasi-thermal spectrum peaking in the optical/UV and rapidly evolving toward redder wavelengths \citep[e.g.][]{Pian17,Drout17}, in overall agreement with kilonova models \citep[e.g.][]{Kasen17,Kasen15}. 
The color evolution of GRB~160821B (Figure~\ref{fig:color})
is consistent with a similar behavior. 
By using a simple power-law model, $F_{\nu}\propto^{-\beta}$ to describe the data, we derive $\beta$=0.70$\pm$0.20 at $T_0 + $2\,hr, rather standard for an afterglow spectrum with $\nu_m<\nu<\nu_c$ \citep{Granot02}.
Data from the first night of observations show a flatter spectrum, $\beta$\,$\approx$0.30, although with larger uncertainties.
The optical/nIR counterpart displays a much redder color, $\beta \approx$1.4--1.8, between 2 and 4 days, and then return to a more typical value of $\beta$=1.1$\pm$0.30.
Evidence of a possible spectral break at $\lambda$\,$\approx$10,000 \AA~is seen in the GTC OSIRIS and CIRCE data at $T_0$+2~d, but no longer visible at later times.
Whereas spectral breaks are observed in many GRB afterglows, the change in spectral slope $\Delta \beta$\,$\gtrsim$1.0 implied by our observation is
hard to reconcile with standard afterglow theory.
 Substantial color variation is often observed during the early ($\lesssim$10$^3$\,s) afterglow phases \citep[e.g.][]{Melandri2017}, 
but is uncommon on longer timescales \citep{Li2018}. 
The observed color evolution in GRB~160821B appears atypical for
a GRB afterglow, and instead consistent with the onset of a kilonova peaking in the optical at $\lesssim$1 d, then rapidly shifting to longer wavelengths. In our last observation,  the intermediate value of the spectral index suggests a significant contribution from the underlying afterglow. This implies that the IR emission started to fade a few days after the burst, as observed in AT2017gfo, and  is therefore shorter lived than the candidate kilonova component in GRB130603B \citep{Tanvir13}. 

No significant emission from the kilonova AT2017gfo was detected at X-ray or radio wavelengths \citep{Troja17,Hallinan17}. 
Based on this evidence, we use the X-ray and radio emission to trace the underlying afterglow component, and compare it to the optical/nIR dataset in order to detect any excess from the kilonova. 
In order to model the afterglow, we consider a standard scenario \citep[e. g.][and references therein]{Zhang04} in which the interaction between the jet and the environment generates two shocks: a highly relativistic forward shock (FS) propagating into the outer medium, and a mildly relativistic reverse shock (RS) traveling backward into the ejecta. 
The shocked electrons are accelerated into a power-law distribution, $N(E) \propto E^{-p}$, and emit their energy via synchrotron radiation.  The resulting broadband spectrum is 
characterized by four quantities: the self-absorption frequency $\nu_a$, the synchrotron frequency $\nu_m$, the cooling frequency $\nu_c$ and the peak flux $F_{pk}$, where we use the subscript FS and RS to distinguish the two spectral components.

Template light curves for the kilonova emission were synthesized
by interpolating over the rest-frame spectra of AT2017gfo, and  stretching times by a factor of (1 + $z$).
We used the spectroscopic data from VLT/X-Shooter \citep{Pian17,Smartt17} and {\it HST} \citep{Troja17} for $t$\,$>$1.5~d,
and photometric measurements from \citet{Drout17,Evans17,Tanvir17,Troja17} at earlier times.

\subsection{Basic Constraints to the Afterglow}\label{ag}

\begin{figure}
\includegraphics[width=0.98\columnwidth]{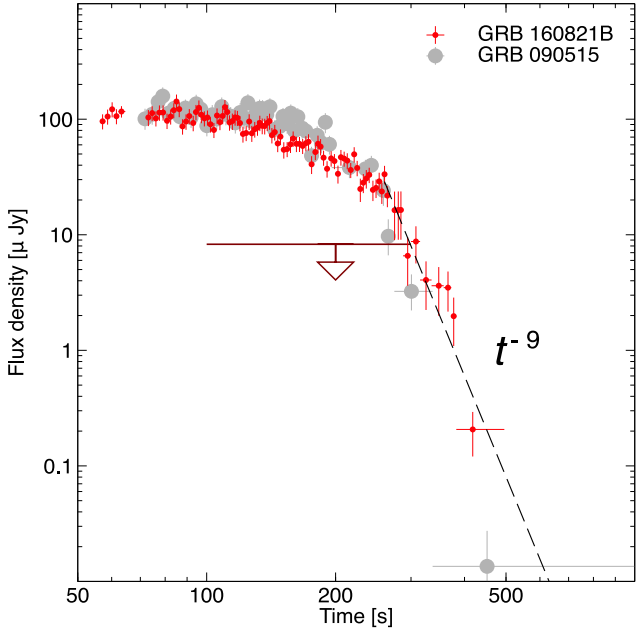}
     \caption{Temporal evolution of the early X-ray afterglow of GRB160821B (red circles) and, for comparison, GRB090515 (gray circles; \citealt{Rowlinson10}). 
     The sharp drop in flux ($\propto t^{-9}$, dashed line) and the deep upper limit from {\it Swift}/UVOT rule out an external shock origin for the observed X-ray emission. }
     \label{fig:xlc}
     \vspace{-0.3cm}
\end{figure}

The early X-ray afterglow displays a bright and rapidly fading light curve (Figure~\ref{fig:xlc}). The measured X-ray flux
decreases from 2$\times$10$^{-10}$\,erg\,cm$^{-2}$\,s$^{-1}$ at 250~s post-burst
to 2$\times$10$^{-12}$\,erg\,cm$^{-2}$\,s$^{-1}$ at 400~s, 
implying a temporal slope  $\alpha$\,$\approx$9, where 
$F_X\propto t^{-\alpha}$.
This sharp decay cannot be reproduced by standard FS models and is generally attributed to the sudden cessation of central engine activity \citep{Troja07,Rowlinson10}. 
Other scenarios, invoking a bright RS emission \citep[e.g.][]{uhm07,vanEerten14}, are not consistent with the simultaneous optical upper limits, nor with the constraints from the radio afterglow data (see below). We therefore attribute this first phase of the afterglow to the continuous activity of the central engine \citep[see also][]{Lu17}, and exclude it from subsequent modeling.

After the first orbit, the X-ray afterglow follows a simple power-law decay with slope $\alpha_1 = 0.84\pm$0.08.  The condition $\beta_X \gtrsim \beta_{\rm opt} \approx 0.7$ is satisfied for
an electron spectral index $p\approx2.3$ when the cooling frequency $\nu_c^{FS}$
lies within or above the X-ray band.  Given its slow temporal evolution, $\nu_c \propto t^{-0.5}$, its passage does not affect the optical/nIR data over the time span of our observations. This additional evidence adds support to the idea that the observed color evolution is not related to the GRB afterglow. 
A comparison to the basic closure relations for FS emission \citep{Zhang04} shows that the temporal slope $\alpha_1$ is shallower than the predicted 
value $3\beta/2$\,$\approx$1.05. We interpret this flattening of the light curve as a viewing angle effect \citep[e.g.][]{Ryan15} due to the contribution from the jet lateral structure.  For this reason, we adopt a Gaussian jet profile, as described in \citet{Troja17} and \citet{Troja18a}, to better model the afterglow evolution.

{\it XMM-Newton} data rule out the presence of an early ($\lesssim$1 d) temporal break, proposed by \citet{Jin18} based on a smaller dataset. They are instead consistent with an uninterrupted power-law decay up to $\approx$4~d (3.5~d rest-frame) and no spectral evolution. The measured spectral index is $\beta = \Gamma -1$ =0.88$\pm$0.24, consistent with the earlier XRT spectrum. 
In the second epoch of XMM observations, the measured flux is three times lower than the predictions based on the simple power-law temporal decay.
 The lower flux is consistent with a steeper power-law decline of slope $\alpha_2 \sim$2.3, 
and may indicate that a temporal break, 
likely a ``jet-break'' \citep{Rhoads99,Sari99}, indeed occurred after 3.5~d.

Radio observations show a fading afterglow since early times. 
For standard FS emission, this may indicate that $\nu_m$ is already below the radio range and that radio, optical, and X-rays belong to the same spectral segment, as observed for GW170817 \citep{DAvanzo2018,Lyman2018,Troja19}. 
However, for GRB160821B the flat radio-to-optical spectral index $\beta_{OR} \approx$0.3 at 4 hr rules out such regime,
and shows that $\nu_m^{FS} \gg $ 6 GHz.  In this case, the fading radio light curve could be explained by an early jet-break (e.g. GRB140903A; \citealt{Troja16}), a flaring episode (e.g. GRB050724A; \citealt{Berger05}), or a RS component from the shock-heated relativistic ejecta  \citep[e.g. GRB051221A;][]{Soderberg06}. 
Simultaneous X-ray data allow us to exclude the first two options,
favoring the RS scenario.
For a short GRB with $T_{90} \approx$0.5~s, the more likely scenario is the thin-shell case leading to a Newtonian RS \citep{Lloyd18,Becerra19}. 
The early radio observations constrain the 
RS peak flux $F_{pk}^{RS}\gtrsim$26 $\mu$Jy and frequency $\nu_{pk}^{RS}\lesssim$6 GHz at 3.5 hrs. Given the predicted evolution of the RS characteristic frequency $\nu_{m}^{RS} \propto t^{-1.5}$ \citep{Kobayashi00}, the RS peak was well below the optical range at 100 s, and therefore did not contribute to the early X-ray emission. 
Optical upper limits, down to $wh\gtrsim$21 mag at $\approx$100 s (Figure~\ref{fig:xlc}), also exclude the presence of a bright RS component at early times. 

Radio observations at later times help constrain the FS peak flux, $F_{pk}^{FS}\lesssim$50 $\mu$Jy, and synchrotron frequency, $\nu_{pk}^{RS}\gtrsim$100~GHz at 1 d. 
For a homogeneous circumburst medium, the radio flux should rise as $t^{1/3}$, and eventually become visible at late times. 
The lack of detection at 10~d and 17~d therefore supports the presence of an earlier  jet-break. 
 
When viewed within the effective core opening angle $\theta_{core}$, that is the viewing angle $\theta_{view} \lesssim \theta_{core}$, the afterglow of a FS with lateral structure is similar to that of a top hat jet viewed on-axis and the jet break time scales roughly as $\theta_{core}^{8/3}$, with deviations as $\theta_{view}$ approaches $\theta_{core}$ \citep{Rossi02}.  
Numerical simulations of initially top-hat jets show that the observed jet-break time scales with these angles as $t_j = 3.5 {\rm~d}\times (1+z)E_{50}^{1/3} n_{-3}^{-1/3} ((\theta_{core} + \theta_{view})/0.2)^{8/3}$, where $E_{50}$ is the on-axis isotropic equivalent energy in units of $10^{50}$\,erg and $n_{-3}$ is the circumburst medium number density in units of $10^{-3}$\,cm$^{-3}$ \citep{vanEerten10}.  Then, using fiducial values $E_{50} = 1$ and $n_{-3}=1$, the observed break in the light curve is consistent with a jet break if $\theta_{core} + \theta_{view} \approx 0.2$~rad.

\begin{figure*}
\includegraphics[width=0.94\columnwidth]{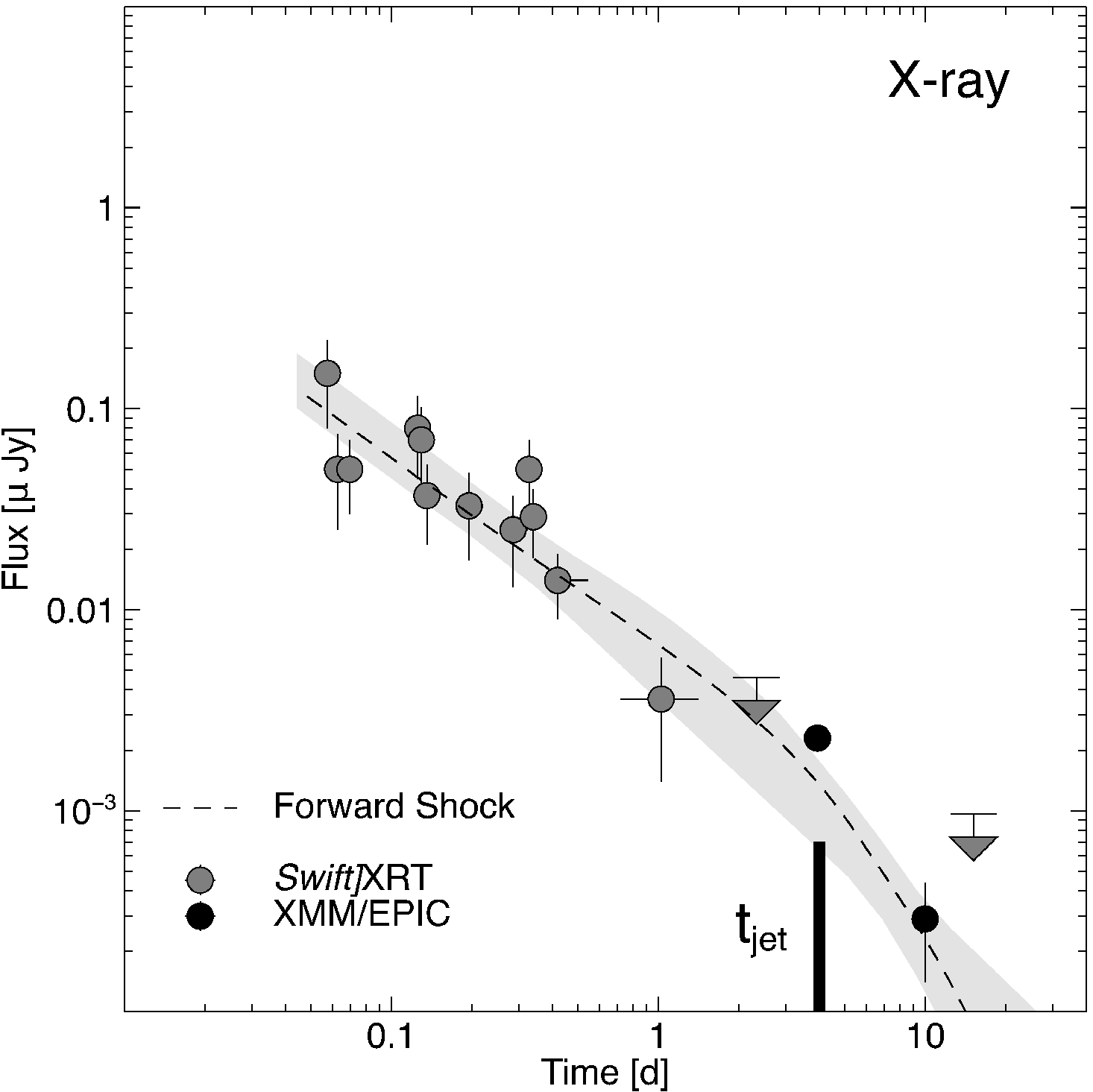}\hspace{0.6cm}
\includegraphics[width=0.94\columnwidth]{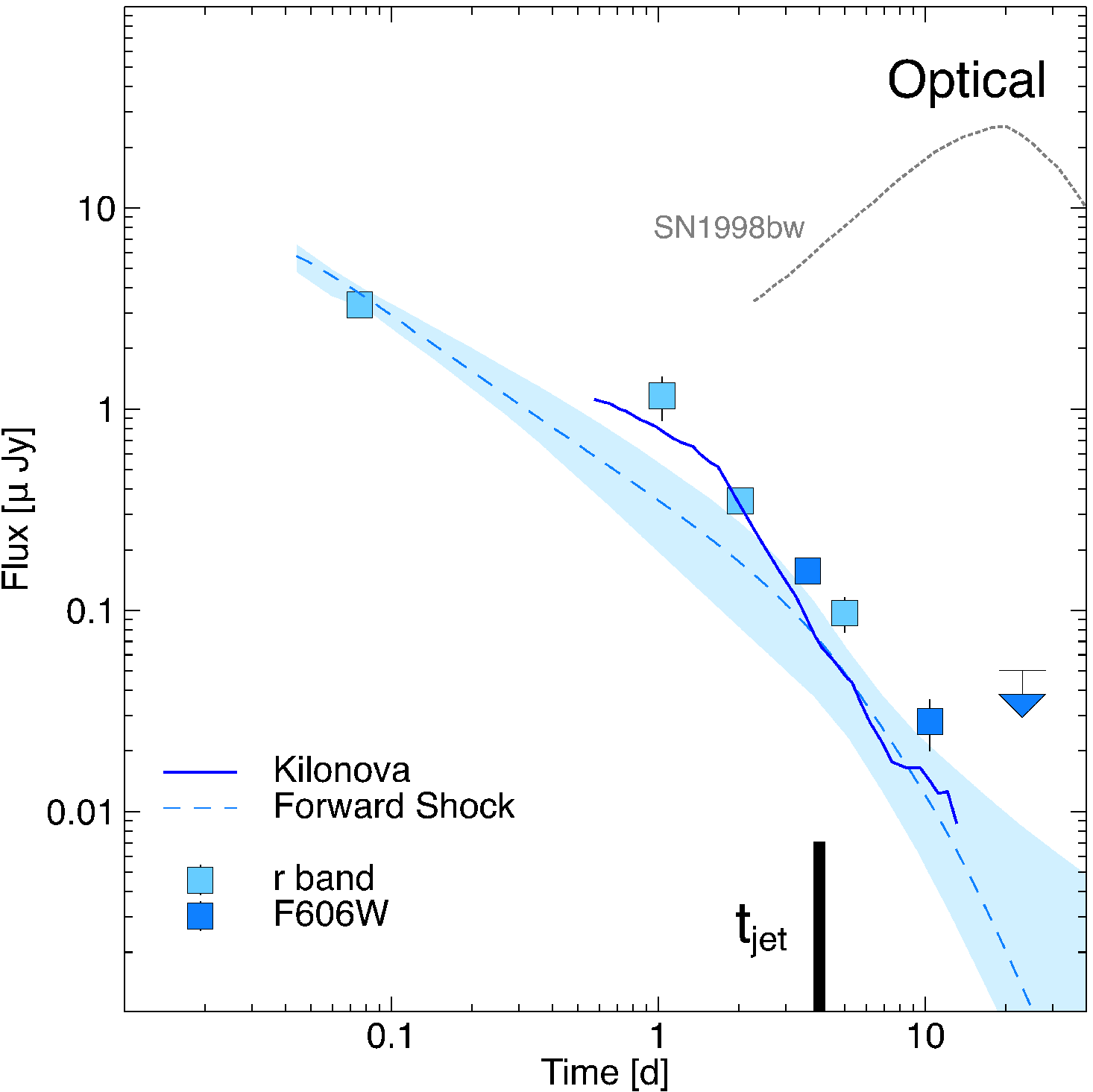}\\
\vspace{0.7cm}
\includegraphics[width=0.94\columnwidth]{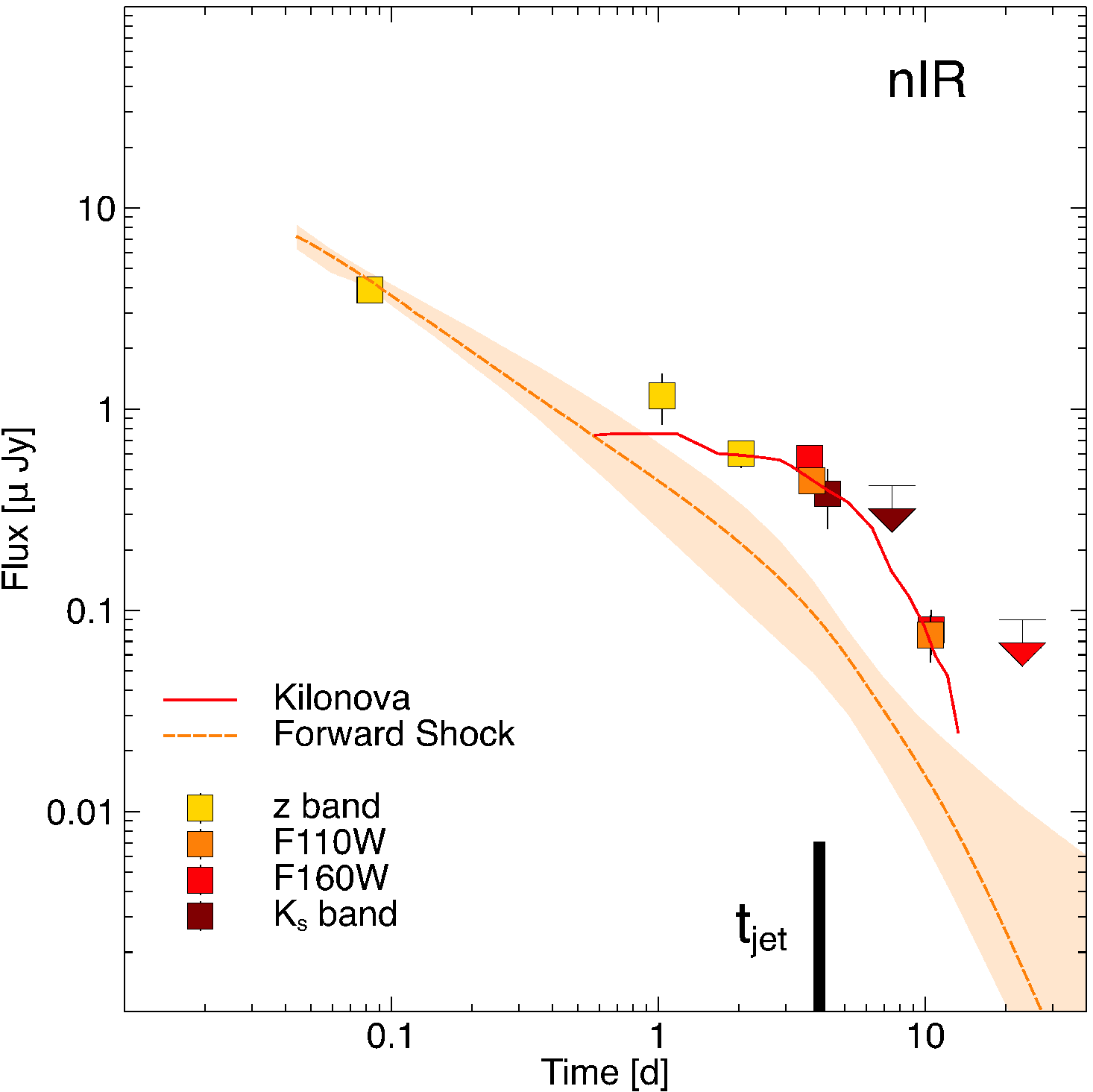}\hspace{0.6cm}
\includegraphics[width=0.94\columnwidth]{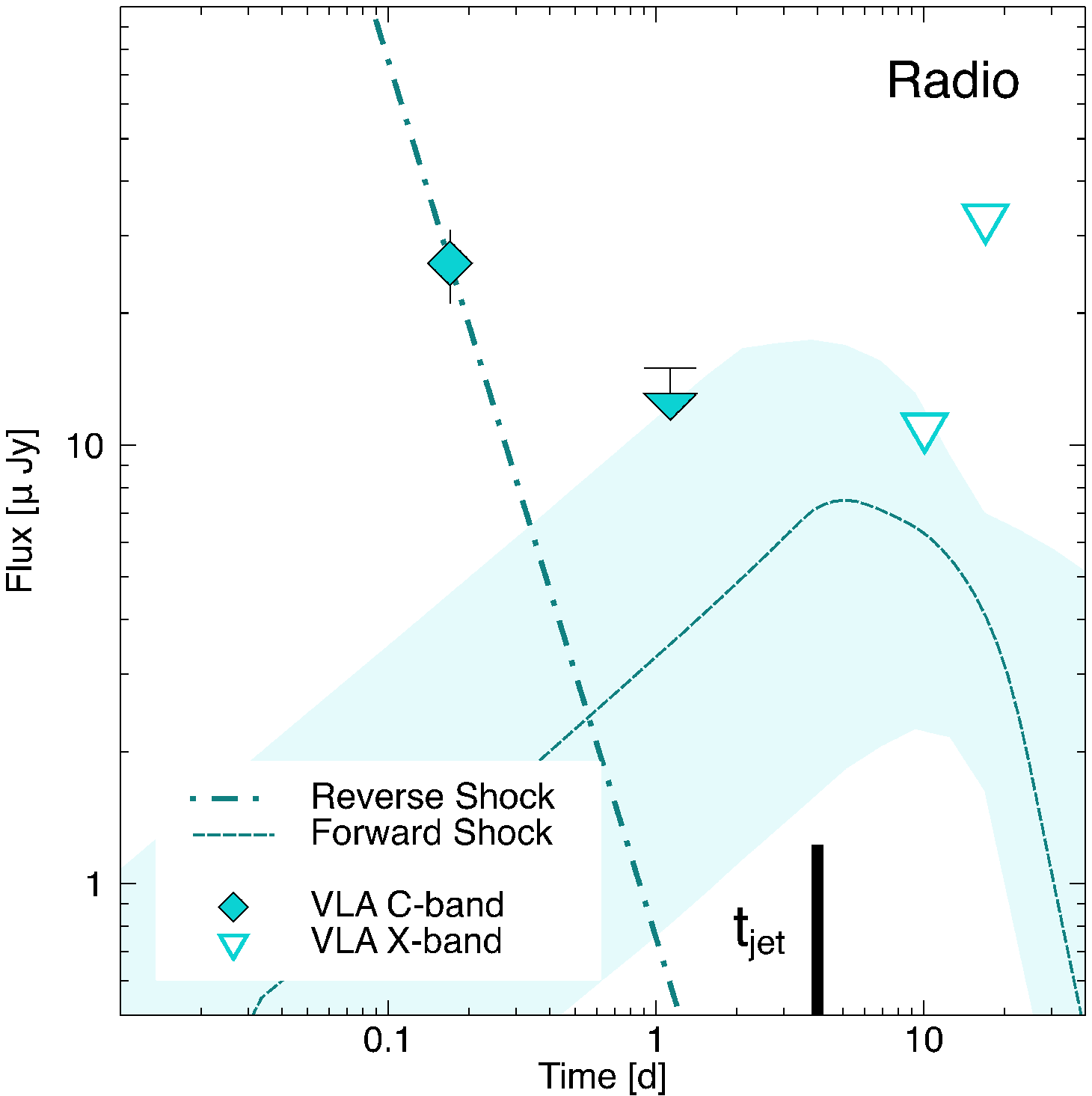}\\
     \caption{Multi-color lightcurves of GRB160821B compared to the
     standard FS (dashed line) + RS (dot-dashed line) scenario. 
     The FS model is the best fit to the broadband dataset,
     whereas the RS is described as a fast fading power-law
     of slope $\approx$2 \citep{Kobayashi00}.
     The jet-break time $t_{\rm jet}$ is shown by the thick vertical line. The shaded areas show the 68\% unceertainty in the model.
     Excess emission at optical and nIR wavelengths is compared with the template kilonova light curves of AT2017gfo (solid line).
     The redshifted optical light curve of SN1998bw 
     (dotted line; \citealt{Galama98}) is also shown for comparison.
     Errors are 1\,$\sigma$, downward triangles are 3\,$\sigma$ upper limits. For plotting purposes,  $r$, $z$ and $K_s$ data were rescaled using the observed colors (Fig.~\ref{fig:color}) in order to match the F606W and F110W filters, respectively.}
     \label{fig:mwlc}
     \vspace{-0.3cm}
\end{figure*}

\begin{figure*}
\includegraphics[width=2.0\columnwidth]{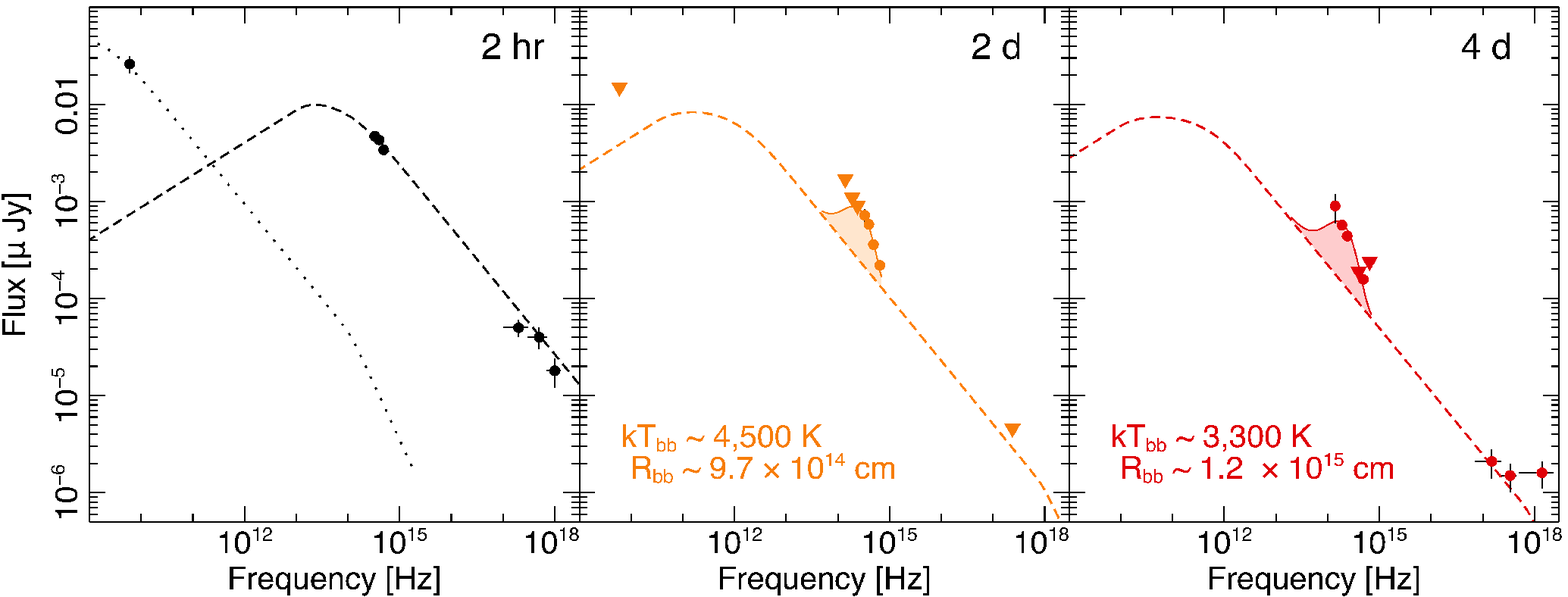}
     \caption{ Broadband spectral energy distribution of GRB160821B.
     Downward triangles are 3~$\sigma$ upper limits. 
     A standard non-thermal afterglow, including a forward shock (dashed line) and a reverse shock (dotted line) component, can reproduce the early data ($T_0$+2~hr). 
     At later times, an additional emission component is visible in the optical/nIR range. It can be described by a black-body spectrum (solid line) with decreasing temperature.  }
     \label{fig:sedkn}
     \vspace{-0.3cm}
\end{figure*}

\subsection{Broadband Afterglow Modeling}
\label{model}

Based on the preliminary constraints discussed in Sect.~\ref{ag}, 
we described the afterglow emission using a standard FS model 
and a structured jet with a Gaussian profile of width $\theta_{core}$. 
We included in the fit the X-ray data, the late radio upper limits, and the first epoch of GTC optical/nIR observations ($\approx$\,$T_0$+0.08\,d).  The rest of the optical/nIR data (from $T_0$+1\,d to $T_0$+10.5\,d; Table~\ref{tab:log}) were treated as upper limits to the afterglow flux. The early radio detection, dominated by RS emission, was also not included. 

Prior to the fit all the data were corrected for Galactic extinction. Given the evidence for negligible intrinsic absorption in the X-ray spectra, and also considered the GRB location in the outskirts of its host galaxy, we assumed $A^{host}_{V}\approx$0. 
A small amount of extinction may affect our estimates of the spectral slope, but not the evident color evolution (Figure~\ref{fig:color}).

We followed the same procedure used in \citet{Troja18a} and \citet{Troja19}, and performed a Bayesian Markov-Chain Monte Carlo (MCMC) model fit to data from synthetic detections generated from our jet model.
Despite the large number of free parameters,  
the FS properties are reasonably well constrained:
\begin{eqnarray*}
\centering
{\rm log~} E_{\rm 0,iso} ~({\rm erg}) & = & 50.4 ~(-0.3, +0.7)  \\
{\rm log~} n ~( {\rm cm}^{-3}) & = &  -2.8  ~(-1.4, +1.1) \\
{\rm log~} \epsilon_e & = &  -0.17  ~(-0.22, +0.12) \\
{\rm log~} \epsilon_B & = & -2.1 ~(-1.0, +1.0)   \\
p & = & 2.31 ~(-0.05, +0.08)  \\
\end{eqnarray*}
\begin{eqnarray*}
\theta_{core} ~({\rm rad}) & =& 0.14 ~(-0.06, +0.36) \\
\theta_{view} ~({\rm rad}) &  = & 0.17 ~(-0.08, +0.25) 
\end{eqnarray*}
where $E_{\rm 0,iso}$ is the isotropic equivalent 
blastwave energy, $n$ the ambient density, 
$\epsilon_e$ and $\epsilon_B$ the shock microphysical parameters,
and $\theta_{view}$ the observer's angle with respect to the jet-axis.
The resulting prompt radiative efficiency is
$\eta_{\gamma} = E_{\gamma,\rm iso} / \left(E_{\gamma,\rm iso} + E_{0,\rm iso} \right)$ = 0.05$^{+0.05}_{-0.04}$. Viewing angle effects are taken into account into our modeling and the derived $E_{0,\rm iso}$, but not in the observed $E_{\gamma,\rm iso}$. Accounting for them would increase its value by a factor $\approx$3, that is $\eta_{\gamma} \approx$0.15.

Our best fit model and its uncertainty are shown in Figure~\ref{fig:mwlc}.
 In this model early emission at $T_0+0.1$\,d is dominated by a small patch around the line of sight with Lorentz factor $\sim 8$. The patch widens as the jet decelerates, through the jet break, until at $T_0+10$d whole jet is in sight and has a Lorentz factor of roughly $1.5$.
Whereas at early times ($< T_0$+0.1\,d)
it provides a good representation of the broadband dataset, 
at later times it does not naturally account for the drastic color change of the optical/nIR data. It therefore underpredicts
the observed emission at nIR and, to a less extent, optical
wavelengths. 
In Figure~\ref{fig:mwlc} we also overplot the template
light curves for AT2017gfo in the $F606W$ and $F110W$ filters,
rescaled to match the observed fluxes. 
The close resemblance between the temporal evolution of the optical /nIR excesses and the kilonova light curves
suggests that they share a common origin. 

 Figure~\ref{fig:sedkn} reports the spectral energy distribution at three different epochs. Whereas at early times ($T_0$+2 hr) the broadband spectrum can be described by an afterglow model, 
at later times emission at optical/nIR wavelengths is brighter and redder
than the model predictions (dashed line). This excess can be modeled with a black-body spectrum of decreasing temperature, from $\approx$4,500~K at 2~d to $\approx$3,300~K at 4~d, and radius $R\approx$10$^{15}$\,cm.

\section{Kilonova properties}

\subsection{Comparison to AT2017gfo and GRB130603B}

\begin{figure}
\includegraphics[width=0.95\columnwidth]{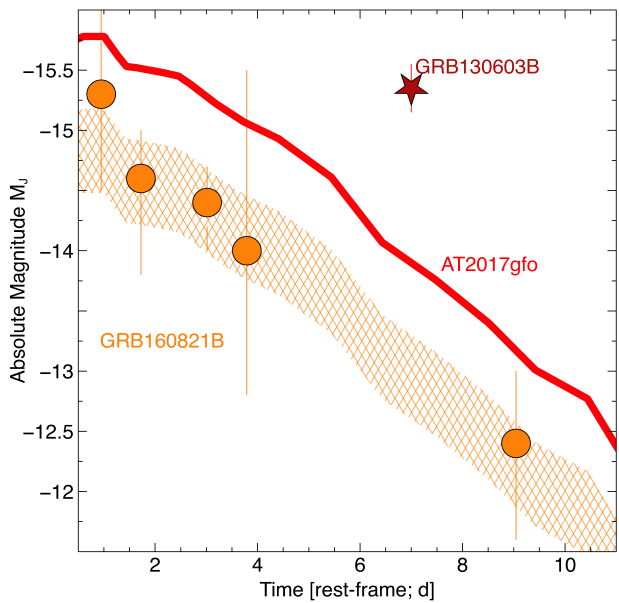}\\
     \caption{Rest-frame nIR light curves for AT2017gfo (red solid line) and the two candidate kilonovae in GRB 130603B (star)
     and GRB 160821B (circles). Error bars are 1\,$\sigma$ statistical uncertainties. The hatched
     area shows the uncertainty in the afterglow subtraction.}
     \label{fig:nirkn}
     \vspace{-0.5cm}
\end{figure}

We interpret the red excess detected in GRB160821B as kilonova emission from fast-moving lanthanide-rich ejecta. 
This is only the second case of a short GRB with a kilonova detection in the nIR, where the emission 
is determined by the heaviest elements ($A$\,$\gtrsim$140). 
In Figure~\ref{fig:nirkn}, we compare its properties to the kilonova AT2017gfo and to the other candidate kilonova in GRB130603B. 
The IR emission of GRB130603B is brighter and longer lived, 
and its properties are not a good match,
as reported in our preliminary analysis of this event \citep{TrojaGCN16}. 
The observed emission resembles more closely the color and temporal evolution of AT2017gfo, although less luminous by $\approx$1 mag.
Our results are consistent with the conclusions of \citet{Jin18},
and provide a better temporal and spectral coverage of the candidate kilonova. 

The complexity of these systems, in particular the 
poorly known nuclear physics involved, 
lead to large uncertainties in the modeling of kilonovae light curves and spectra \citep[e.g.][]{Rosswog17}. 
We do not attempt here a systematic comparison to the
various models presented in the literature,
but estimate the basic explosion parameters based on the scaling relations of \citet{Grossman14}. 
The observed nIR luminosity, $L_{\rm nIR}$\,$\approx$2$\times$10$^{39}$\,erg s$^{-1}$,
and timescales, $t_{\rm pk}$\,$\lesssim$\,3\,d, 
imply a low ejecta mass $M_{ej}$\,$\lesssim$0.006 $M_{\odot}$ and high velocity
$v_{ej}$\,$\gtrsim$0.05$c$ for an opacity $\kappa$\,$\approx10$\,g\,cm$^{-3}$. This result agrees well with the constraints $v \approx R_{bb}/ t \approx$0.1-0.2\,$c$, derived from simple black body fits (Figure~\ref{fig:sedkn}).
The ejecta mass is substantially lower than the values inferred for other GRB kilonovae \citep{Jin16,Yang15,Tanvir13}, and comfortably within
the range of dynamical ejecta from double NS mergers. 

Our analysis also finds evidence for an early blue excess, although with larger uncertainties. It is suggestive that the luminosity and timescale
of this blue component are consistent with the early optical emission in AT2017gfo (Figure~\ref{fig:mwlc}, top right panel).
The blue color and early onset require a larger mass ($M_{ej}$\,$\approx$0.01 $M_{\odot}$) of lanthanide-poor material, 
produced, for example, by the merger remnant.

\subsection{Effects of a long-lived NS}
The merger of two NSs can lead either to a stellar-mass BH or to a hypermassive highly magnetized NS \citep{Giacomazzo13,Piro17}. 
The latter is thought to significantly affect both the kilonova colors and the afterglow evolution through its continuous energy injection and strong neutrino irradiation \citep{Kasen15,Gao15,Lippuner17,Radice18}. 
Indeed, the red colors of GW170817/AT2017gfo and its smooth afterglow light curve, mostly consistent with a standard FS emission, were used to exclude a wide range of possible NS configurations \citep{Margalit17,Ai18}. Only a short-lived proto-magnetar or a long-lived NS with a weak poloidal field could be consistent with the 
electromagnetic properties of GW170817 \citep{Li18,Metzger18,Yu18,Piro19}. 

GRB160821B  allows us to study the link between the GRB central engine and the kilonova properties. Its X-ray afterglow shows evidence of a long-lasting engine activity. In particular, the early X-ray emission shows a phase of nearly constant flux followed by a sharp decay (Figure~\ref{fig:xlc}) often interpreted as a signature of a proto-magnetar \citep{Fan06,Troja07,Liang07,Lyons10,Rowlinson10,DallOsso11}. The sudden cessation of X-ray emission at $t\approx200$~s may be caused by the NS collapse into a BH \citep{Troja07}. 

Within this framework, the observed plateau luminosity, $L_X$\,$\approx$5$\times$10$^{47}$\,erg\,s$^{-1}$,  and its lifetime, $T$\,$\approx$200~s, can be used to infer the magnetar properties \citep{Zhang01}. 
For an isotropic emission, \citet{Lu17} derived 
an initial spin period $P_0 \approx$9~ms and a dipolar magnetic field $B_0 \approx$3$\times$10$^{16}$~G, in overall agreement with the assumption of a magnetar which suffered from dominant energy loss via gravitational wave radiation \citep{Fan2013}.
Beaming would substantially affect these results, and yield unphysical values of magnetic field and period. This indicates that, if the power source of the early X-ray emission is a newborn spindown NS, then the magnetar-driven outflow is nearly isotropic, in agreement with the recent claim of magnetar-driven fast X-ray transients \citep{Xue19}.

Prolonged irradiation from the NS remnant will affect the ejecta composition and velocity profile, resulting in a bluer and short-lived kilonova emission.
However, the color evolution of GRB160821B supports the presence of a red kilonova evolving on timescales similar to AT2017gfo, although less luminous by a factor of $\approx$2--5. 
While the lower IR luminosity could be an effect of the long-lived NS, the red excess shows that, despite it, a good amount of lanthanide-rich material was formed and released into the ambient medium, e.g. by tidally stripped matter ejected before the merger \citep{Korobkin12}.

\section{Environment}

 Late-time {\it HST} observations (Table~1) place deep upper limits on any galaxy underlying the GRB position. Such faint host
galaxies are rarely observed in short GRBs \citep[e.g.][]{Piranomonte2008} and imply, at least, a moderately high redshift $z\gtrsim$1.  Similar distance scales are disfavored by the presence of a kilonova, for which luminosities $>>$10$^{43}$\,erg would lead to implausible properties of the merger ejecta.
We consider more natural to link GRB160821B with the nearby bright galaxy (Figure~\ref{fig:one}), whose spectrum and disturbed morphology are consistent with a star-forming spiral galaxy, possibly undergoing a merger phase. At the measured offset of 5.7\arcsec, the probability of a chance alignment between the GRB and the galaxy is $\approx$2\%. However, we find that the overall galaxy properties suggest an environment typical of short GRBs, thus strenghtening the case for a physical association.

The galaxy's absolute B-band magnitude is $M_B \approx$ -19.9 AB, 
approximately 0.7$L^{*}_B$ for a late-type galaxy \citep[Type~3+4;][]{Zucca09},
and its colors are consistent with those of an irregular galaxy \citep{Fukugita95}.
It is detected in the WISE bands with $W1$(3.4 $\mu$m) = 20.52$\pm$0.10 and $W2$(4.6 $\mu$m) = 21.0 $\pm$ 0.3 AB mag, from which we estimate a stellar mass
log ($M/M_{\odot}$) $\approx$ 8.5 \citep{Wen13}, at the lower end of the short GRB distribution \citep{Berger14}.

We used standard emission line diagnostics to infer the average properties of the putative host. The prominent nebular emission lines are indicative of on-going star formation. The [OII] line luminosity gives a star formation rate  SFR([O II]) $\gtrsim$\,1.5\,$M_{\odot}$\,yr$^{-1}$ \citep{Kennicutt98},
consistent with the estimate  SFR(FUV) $\approx$\,1.2\,$M_{\odot}$\,yr$^{-1}$ derived from the UV luminosity
$M_{w2}$\,$\approx$-18.5 AB mag \citep{Hao2011}. 
Line ratios, log ([OIII]/$H\beta$)$\approx$0.4 and log ([OIII]/[OII])$\approx$-0.1, are substantially lower than in long GRB host galaxies, and within the range of short GRB hosts \citep{Levesque07}.
The solution based on the $R_{23}$ \citep{Pagel79} and $O_{32}$ indicators is degenerate,
and we can only constrain the upper metallicity branch \citep{Kobulnicky04}, for which we find
12 + log(O/H) $\approx$ 8.7, consistent with a solar metallicity \citep{Asplund09}.

 The projected physical offset is $\approx$16.4~kpc from the galaxy's center, at the higher end of the offset distribution for short GRBs \citep{Troja08,Fong13,Tunnicliffe14}. 
As only a small fraction of stars is expelled during galaxy 
interactions \citep[e.g.][]{Behroozi13}, the observed offset is likely the result of an intrinsic kick imparted to the progenitor upon birth. 
For a stellar age of $\approx$300 Myr
and a velocity dispersion  $v_{disp} \approx$120\,km\,s$^{-1}$, a minimum kick velocity $v_{kick} \gtrsim$80\,km\,s$^{-1}$ would be required to reproduce the observed distance. However, a more detailed treatment, accounting for the galaxy's potential as well as  the possibility of multiple orbits around the galaxy, 
finds that natal kicks larger than 150\,km\,s$^{-1}$ are needed to explain the large offsets of some  short GRBs \citep{Behroozi14}.

 An alternative explanation for short GRBs with large offsets was discussed by \citet{Salvaterra2010}, who investigated compact binaries dynamically formed in globular clusters. At the distance of the candidate host, {\it HST} upper limits 
can only exclude the presence of luminous cluster systems, such as those found in M87, but are otherwise unconstraining. The range of circumburst densities from afterglow modeling (Sect. 3.2) is also compatible with an origin in a globular cluster \citep{Freire2001}.


\section{Conclusions}

GRB160821B is a nearby ($z$\,$\sim$\,0.1613) short GRB with bright X-ray, radio and optical/nIR counterparts.  The X-ray emission shows evidence for continued energy injection from a long-lived central engine, active up to $\approx$200~s after the burst. At later times, the X-ray afterglow is consistent with standard forward shock emission, whereas the radio signal was likely dominated by a weak reverse shock.
Optical/nIR observations show a clear evolution toward red colors, consistent with the onset of a lanthanide-rich kilonova similar to AT2017gfo.  The identification of this additional component is challenging due to the contamination from the underlying bright afterglow, and required an uncommonly rich dataset to be disentangled. 
Within the sample of short GRBs, this is only the second kilonova detected in the nIR. Its low luminosity implies a low mass of lanthanide-rich ejecta, possibly as an effect of a long-lived NS remnant.

\section{Acknowledgements}

This work is dedicated to the memory of John K. Cannizzo, a friend and colleague with whom we shared many stimulating conversations about kilonovae and short GRBs.

We thank the anonymous referee for his/her insightful and constructive comments. We gratefully thank R. Clavero and S. Dichiara for assistance,
B. Metzger and J. Barnes for help in the preliminary modeling.
We acknowledge the use of public data from the Swift data archive.
This work made use of data supplied by the UK Swift Science Data Centre at the University of Leicester. These results also made use of Lowell Observatory's Discovery Channel Telescope. Lowell operates the DCT in partnership with Boston University, Northern Arizona University, the University of Maryland, and the University of Toledo. Partial support of the DCT was provided by Discovery Communications. LMI was built by Lowell Observatory using funds from the National Science Foundation (AST-1005313). 
The work is partly based on the observations made with the Gran Telescopio Canarias (GTC), installed in the Spanish Observatorio del Roque de los Muchachos of the Instituto de Astrofisica de Canarias, in the island of La Palma. 

ET acknowledges financial support provided by the National Aeronautics and Space Administration through HST-GO14357, HST-GO14087, HST-GO14607 and HST-GO14850 grants from the Space Telescope Science Institute, operated by the Association of Universities for Research in Astronomy, Incorporated. 
AJCT, YDH e IM acknowledge financial support from the State Agency for Research of the Spanish MCIU through the "Center of Excellence Severo Ochoa" award for the Instituto de Astrof\'isica de Andaluc\'ia (SEV-2017-0709). 
The development of 
CIRCE at GTC was supported by the University of Florida and the National 
Science Foundation (grant AST-0352664), in collaboration with IUCAA.
RSR acknowledges support by Italian Space Agency (ASI) through the Contract n. 2015-046-R.0 and by AHEAD the European Union Horizon 2020 Programme under the AHEAD project (grant agreement n. 654215). YDH also acknowledges the support by the program of China Scholarships Council (CSC) under the Grant No.201406660015.
JBG acknowledges the support of the Viera y Clavijo program funded by ACIISI and ULL.
GN and AT acknowledge funding in the framework of the project ULTraS (ASI--INAF contract N. 2017-14-H.0).  GR acknowledges the support of the University of Maryland through the Joint Space Science Institute Prize Postdoctoral Fellowship.

\bibliographystyle{aa}


\label{lastpage}
\end{document}